\documentclass[12pt]{iopart}
\usepackage[dvips]{graphicx}
\begin{document}
\title[Finite-dimensional representation of the quadratic algebra]{Finite-dimensional representation of the quadratic algebra of a generalized coagulation-decoagulation model}%
\author{Farhad H. Jafarpour\footnote{Corresponding author's e-mail: farhad@ipm.ir}\, and Ali Aghamohammadi}
\address{Bu-Ali Sina University, Physics Department, Hamadan, Iran}
\begin{abstract}
The steady-state of a generalized coagulation-decoagulation model on a one-dimensional lattice with reflecting boundaries is studied using a matrix-product approach. It is shown that the quadratic algebra of the model has a four-dimensional representation provided that some constraints on the microscopic reaction rates are fulfilled.
The dynamics of a product shock measure with two shock fronts, generated by the Hamiltonian of this model, is also studied. It turns out that the shock fronts move on the lattice as two simple random walkers which repel each other provided that the same constraints on the microscopic reaction rates are satisfied.
\end{abstract}
\maketitle
%%%%%%%%%%%%%%%%%%%%%%%%%%%%%%%%%%%%%%%%%%%%%%%%%%%%%%%%%%%%%%%%%%%%%%%%%%%%%%%%%%%%%%%%%%%%%%%%%%%%%%%%%%%%%%%%%%%%%%%
\section{Introduction}
The study of microscopic structure and dynamics of traveling wave solutions in multi-species one-dimensional stochastic systems has attracted people's attention in this field considerably in recent years \cite{FER}-\cite{PS}. For instance, the microscopic dynamics of shocks are studied for three families of single-species one-dimensional reaction-diffusion systems with open boundaries and nearest-neighbors interactions which include the Partially Asymmetric Simple Exclusion Process (PASEP), the Branching-Coalescing Random Walk (BCRW) and the Asymmetric Kawasaki-Glauber Process (AKGP) \cite{KJS}. It has been shown that in all three systems the time evolution of a product shock measure with a single shock front is equivalent to that of a simple random walker on a finite lattice with homogeneous hopping rates in the bulk and special reflection rates at the boundaries, provided that some constraints on the microscopic reaction rates are fulfilled. The steady-states of these three systems can be essentially written as a linear superposition of such product shock measures. On the other hand, the steady-states of these systems can be obtained using the matrix-product formulation \cite{J2} in which the steady-state weights are written in terms of the product of non-commuting operators which satisfy a quadratic algebra (for a recent review of this approach see \cite{BE}). Surprisingly, it has been found that the conditions under which these operators have two-dimensional matrix representations are exactly those for a product shock measure to have a simple random walk dynamics in these systems \cite{J2}.\\
It has been shown in \cite{JM2} that the existence of a two-dimensional representation for the quadratic algebra of a multi-species one-dimensional reaction-diffusion system with open boundaries and nearest-neighbors interactions implies that the steady-state of the system can be written in terms of a linear superposition of product shock measures. It seems that the same is true for the systems whose quadratic algebras have higher-dimensional matrix representations. For instance, for the Totally Asymmetric Simple Exclusion Process with open boundaries it is known that the quadratic algebra has an infinite-dimensional matrix representation. It has been shown that this is associated with the fact that the steady-state of this system can be written as a linear superposition of product shock measures with infinite number of shocks \cite{BS}.\\
The microscopic dynamics of shock fronts has also been studied in driven-diffusive systems with more than a single species of particles and also in these systems with next-nearest-neighbor interactions \cite{RS}-\cite{PS}.\\
Despite of these efforts it is still generally unclear whether the existence of a finite-dimensional matrix representation of the quadratic algebra is related to the fact that the steady-state can be expressed as a superposition of product shock measures. In this paper we consider a generalized one-dimensional single-species coagulation-decoagulation model with reflecting boundaries as a new example. We believe that the study of this exactly solvable model provides us with another piece of evidence for the existence of such relation and definitely sheds more light on the unknown aspects of this problem. We will show that a product shock measure with two shock fronts which have simple random walk dynamics can evolve in this system provided that the microscopic reaction rates satisfy some constraints. This will enable us to construct the steady-state of the system simply by considering a linear superposition of such measures; however, we will not follow this approach here. Instead we will show that under the {\em same constraints} the quadratic algebra of this model has a four-dimensional representation.\\
Our paper is organized as follows: In section (2) we will explain the mathematical preliminaries and introduce the model. In section (3) we will study the dynamics of a product shock measure with two shock fronts and investigate the conditions (by imposing some constraints on the microscopic reaction rates of the model) under which it has a simple time evolution similar to that of two random walkers moving on a one-dimensional lattice while reflecting from the boundaries. The fourth section will be devoted to the investigation of whether a finite-dimensional representation exists for the quadratic algebra of the model under the same conditions. In the last section we will discuss the summery of results.
%%%%%%%%%%%%%%%%%%%%%%%%%%%%%%%%%%%%%%%%%%%%%%%%%%%%%%%%%%%%%%%%%%%%%%%%%%%%%%%%%%%%%%%%%%%%%%%%%%%%%%%%%%%%%%%%%%%%%%%
\section{The model}
We define $\vert P(t)\rangle$ as the probability vector of a Markovian interacting particle system at the time $t$. The time evolution of this vector is governed by the master equation which can be written as a Schr\"odinger equation in imaginary time as follows:
\begin{equation}
\frac{d}{dt}\vert P(t) \rangle=H \vert P(t) \rangle
\end{equation}
in which $H$ is an stochastic Hamiltonian. The matrix elements of the Hamiltonian are the transition rates between different configurations. For a single-species system with nearest-neighbors interactions defined on a one-dimensional lattice of length $L$ with reflecting boundaries the Hamiltonian $H$ is of the following form:
\begin{equation}
\label{H1}
H=\sum_{k=1}^{L-1}h_{k,k+1}
\end{equation}
in which:
$$
h_{k,k+1}={\mathcal I}^{\otimes (k-1)}\otimes h \otimes {\mathcal I}^{\otimes (L-k-1)}
$$
where ${\mathcal I}$ is a $2 \times 2$ identity matrix and $h$ is a $4
\times 4$ matrix for the bulk interactions. In the basis $(00,01,10,11)$ the Hamiltonian $h$ can be written as:
\begin{equation}
\begin{array}{c}
\label{H2}
h = \left( \begin{array}{cccc}
\omega_{11}& \omega_{12} & \omega_{13} & \omega_{14} \\
\omega_{21} & \omega_{22} & \omega_{23} & \omega_{24} \\
\omega_{31} & \omega_{32} & \omega_{33} & \omega_{34} \\
\omega_{41} & \omega_{42} & \omega_{43} & \omega_{44} \end{array} \right)
\end{array}
\end{equation}
in which we have defined the basis vectors:
\begin{equation}
 \vert 1\rangle =
  \left(\begin{array}{c}
    0 \\ 1
  \end{array}\right), \quad
  \vert 0\rangle =
  \left(\begin{array}{c}
    1 \\ 0
  \end{array}\right)
\end{equation}
associated with the presence of a particle and a hole in each lattice site respectively. Requiring the conservation of probability, the sum of the elements in each column should be zero; therefore, one has $\omega_{ii}=-\sum_{j\neq i}\omega_{ji}$ for $i,j=1,\cdots,4$.\\
In our generalized coagulation-decoagulation model the non-zero reaction rates belong to the following reactions:
\begin{equation}
\label{R1}
\begin{array}{ll}
\emptyset+A \rightarrow A+\emptyset & \mbox{with rate} \; \; \omega_{32} \\
A+\emptyset \rightarrow \emptyset+A & \mbox{with rate} \; \; \omega_{23} \\
A+A \rightarrow A+\emptyset & \mbox{with rate} \; \; \omega_{34} \\
A+A \rightarrow \emptyset+A & \mbox{with rate} \; \; \omega_{24} \\
\emptyset+A \rightarrow A+A & \mbox{with rate} \; \; \omega_{42} \\
A+\emptyset \rightarrow A+A & \mbox{with rate} \; \; \omega_{43} \\
\end{array}
\end{equation}
in which $A$ and $\emptyset$ stand for the presence of a particle and a hole in each lattice site respectively. This model has already been studied in related literatures in a couple of special cases. For instance consider the case:
\begin{equation}
\label{R2}
\omega_{24}=\omega_{23}=q^{-1}\; , \; \omega_{34}=\omega_{32}=q \;,\;\omega_{42}=\Delta q \;,\;\omega_{43}=\Delta q^{-1}.
\end{equation}
The complete spectrum of the Hamiltonian in this case has been obtained exactly \cite{HKP}. The steady-state of the system has also been studied using both the method of empty intervals \cite{HKP} and the matrix-product approach \cite{HSP}. It turns out that the phase diagram of the model in this case has two different phases: a low-density phase and a high-density phase which are separated by a coexistence line. The particle concentration on the lattice in both phases have exponential behaviors with three different length scales while on the coexistence line it changes linearly in the bulk of the lattice and grows exponentially near one of the boundaries. By using the matrix product approach it has been shown that for the special tuning of the parameters given in (\ref{R2}) the quadratic algebra of the model has a finite-dimensional matrix representation \cite{HSP}. Since the system has three different length scales and that these length scales are determined by the eigenvalues of these matrices \cite{HKP}, the authors in \cite{HSP} have found that the minimum dimensionality of these matrices should be four.\\
The most general case (\ref{R1}) has already been studied on a finite lattice with open boundaries in which the particles can enter into the system or leave it from the boundaries \cite{KJS} and also on a lattice with periodic boundary conditions \cite{AA1}. Note that the microscopic reaction rates in our generalized model are exactly those of the BCRW studied in \cite{KJS}. As we mentioned before, it is known that traveling wave solutions exist for the BCRW with open boundaries under some constraints on the microscopic reaction rates. In \cite{J2} the author has shown that in the open boundaries case the steady-state can be written as a matrix-product state with two-dimensional matrix representation given that the same constraints are satisfied. The authors in \cite{AA2} have studied the dynamical phase transition in this system when considered on an infinite lattice.
%%%%%%%%%%%%%%%%%%%%%%%%%%%%%%%%%%%%%%%%%%%%%%%%%%%%%%%%%%%%%%%%%%%%%%%%%%%%%%%%%%%%%%%%%%%%%%%%%%%%%%%%%%%%%%%%%%%%%%%
\section{Dynamics of a product shock measure with two shock fronts}
In what follows we consider the most general reaction rates (\ref{R1}) and study the time evolution of a product shock measure with two shock fronts defined as:
\begin{equation}
\label{P}
\begin{array}{c}
  \vert P_{m,n} \rangle =
  \left(\begin{array}{c}
    1 \\ 0
  \end{array}\right)^{\otimes m} \otimes
  \left(\begin{array}{c}
    1-\rho \\ \rho
  \end{array}\right)^{\otimes n-m-1}
\otimes
  \left(\begin{array}{c}
    1 \\ 0
\end{array}\right)^{\otimes L-n+1} \\ \\ \mbox{for} \quad 0 \leq m \leq n-1 \quad \mbox{and} \quad  1 \leq n \leq L+1
\end{array}
\end{equation}
on a lattice of length $L$ with reflecting boundaries. We define two auxiliary sites $0$ and $L+1$ for convenience. In Figure 1 a simple sketch of a double-shock structure is given. We investigate the conditions under which the time evolution equations for (\ref{P}) are simply those of two random walkers at sites $m$ and $n$ moving on a finite lattice. It turns out that by imposing two constraints on the microscopic reaction rates one finds the appropriate answer. These constraints are found to be:
\begin{equation}
\label{Con}
\frac{\omega_{24}+\omega_{34}}{\omega_{42}+\omega_{43}}=\frac{\omega_{23}}{\omega_{43}}=\frac{\omega_{32}}{\omega_{42}}.
\end{equation}
provided that the height of the shocks is:
$$
\rho=\frac{\omega_{42}+\omega_{43}}{\omega_{42}+\omega_{43}+\omega_{24}+\omega_{34}}.
$$
One should note that these constraints are exactly those obtained in \cite{AA1} for the model with periodic boundary conditions.
\begin{figure}
\centering
\includegraphics[height=3cm]{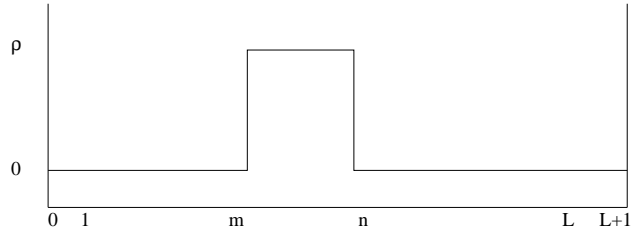}
\caption{A simple sketch of a double-shock measure. The shock
positions are defined at the sites $m$ and $n$.}
\end{figure}
Under these constraints the time evolution equations for $\vert P_{m,n} \rangle$ take the following form:
\begin{equation}
\label{TE}
\begin{array}{l}
H \vert P_{m,n} \rangle = \delta_{1r} \vert P_{m+1,n} \rangle + \delta_{1l} \vert P_{m-1,n} \rangle
+\delta_{2r}\vert P_{m,n+1} \rangle+\delta_{2l} \vert P_{m,n-1}\rangle -\\(\delta_{1r}+\delta_{1l}+\delta_{2r}+\delta_{2l})\vert P_{m,n} \rangle \quad \mbox{for} \quad
m=1,\cdots,L-2 \quad \mbox{and} \\ n=m+2,\cdots,L \\ \\
H \vert P_{0,n} \rangle=-\bar\delta \vert P_{1,n} \rangle+\delta_{2r}
\vert P_{0,n+1} \rangle+\delta_{2l} \vert P_{0,n-1} \rangle
-\\(-\bar\delta+\delta_{2r}+\delta_{2l})\vert P_{0,n} \rangle \quad \mbox{for} \quad n=2,\cdots,L\\ \\
H \vert P_{m,L+1} \rangle=\delta_{1r} \vert P_{m+1,L+1} \rangle+\delta_{1l} \vert
P_{m-1,L+1} \rangle+\bar\delta\vert P_{m,L} \rangle- \\
(\delta_{1r}+\delta_{1l} +\bar\delta)\vert P_{m,L+1} \rangle \quad \mbox{for} \quad m=1,\cdots,L-1\\ \\
H\vert P_{0,L+1} \rangle=-\bar\delta\vert P_{1,L+1} \rangle+\bar\delta\vert P_{0,L} \rangle\\ \\
H\vert P_{m,m+1} \rangle=0 \quad \mbox{for} \quad m=0,\cdots,L
\end{array}
\end{equation}
in which the bulk hopping rates for the shock fronts (random walkers) are:
\begin{equation}
\label{Ref}
\begin{array}{l}
\delta_{1l}=\frac{\omega_{42}}{\rho}, \\
\delta_{1r}=\omega_{43}\frac{(1-\rho)^2}{\rho}+\omega_{24}\rho, \\
\delta_{2l}=\omega_{42}\frac{(1-\rho)^2}{\rho}+\omega_{34}\rho, \\
\delta_{2r}=\frac{\omega_{43}}{\rho}
\end{array}
\end{equation}
and also the reflection rate from the boundaries is found to be:
\begin{equation}
\bar\delta=\delta_{2l}-\frac{\delta_{1r}+\delta_{2l}}{\delta_{1l}+\delta_{2r}}\delta_{2r}.
\end{equation}
Note that the original model has six independent parameters while by imposing the constraints (\ref{Con})
only four independent parameters are remained. \\
The last equation in (\ref{TE}) implies that an empty lattice is a trivial steady-state of the model. It should be mentioned that the random walkers repel each other so that their positions cannot get closer than a single lattice site. This has already been observed in \cite{JM3} for the special choice of parameters given in (\ref{R2}). For the present case, as in \cite{JM3}, one can simply define a new measure $\vert {\tilde P}_{m,n} \rangle$ as:
\begin{equation}
\begin{array}{c}
  \vert {\tilde P}_{m,n} \rangle =
  \left(\begin{array}{c}
    1 \\ 0
  \end{array}\right)^{\otimes m} \otimes
  \left(\begin{array}{c}
    0 \\ 1
  \end{array}\right)^{\otimes n-m-1}
\otimes
  \left(\begin{array}{c}
    1 \\ 0
  \end{array}\right)^{\otimes L-n+1}
\\ \\ \mbox{for} \quad 0 \leq m \leq n-1 \quad \mbox{and} \quad  1 \leq n \leq L+1
\end{array}
\end{equation}
and investigate its dynamics to see that the shock positions never meet each other at two consecutive sites. The reason for this goes back to the fact that the dynamical rules (\ref{R1}) do not allow an empty system to be generated from a system with particles. In fact the system has two different steady-states. The trivial steady-state, as we mentioned above, is the one without any particles. The other one, which is a nontrivial steady-state, contains some particles. If we start with an empty lattice, it always remains empty. In contrast, if we start with a partially-filled lattice, it will reach to an steady-state which contains particles. It can be shown that the nontrivial steady-state of the system can be written as a linear superposition of $\vert P_{m,n} \rangle$'s. This will be discussed and presented elsewhere. In the next section we are going to calculate the steady-state of the system using the matrix-product approach. The question is that whether or not a finite-dimensional matrix representation exists for the quadratic algebra of this model under the constraints (\ref{Con}).
%%%%%%%%%%%%%%%%%%%%%%%%%%%%%%%%%%%%%%%%%%%%%%%%%%%%%%%%%%%%%%%%%%%%%%%%%%%%%%%%%%%%%%%%%%%%%%%%%%%%%%%%%%%%%%%%%%%%%%%
\section{Matrix-product steady-state}
According to the matrix product formalism the stationary probability distribution $P( \tau )$ of any configuration     $\tau=\{ \tau_i \vert i=1,\cdots,L\}$ is assumed to be of the form:
\begin{equation}
\label{Weigth}
P(\tau)=\frac{1}{Z_{L}} \langle W \vert
\prod_{i=1}^{L}(\tau_i D+(1-\tau_i)E)\vert V \rangle
\end{equation}
in which the occupation number $\tau_i$ is defined as follows: $\tau_i=0$ if the site $i$ is empty and $\tau_i=1$ if it is occupied by a particle. The factor $Z_L$ in (\ref{Weigth}) is a normalization factor and called the partition function of the system and can be easily calculated using the normalization condition $\sum_{\{ \tau_i \}}P(\tau)=1$. For a system described by the Hamiltonian of type (\ref{H1}), the operators $D$ and $E$, which stand for the presence of a particle and an empty site at each lattice site respectively, beside the vectors $\langle W \vert$ and $\vert V \rangle$ should satisfy the following quadratic algebra:
\begin{equation}
\begin{array}{c}
\label{MPA}
h  \left[ \left( \begin{array}{c} E \\
D \end{array} \right) \otimes
\left( \begin{array}{c} E \\
D \end{array} \right) \right]=
\left( \begin{array}{c} \bar{E} \\
 \bar{D} \end{array} \right) \otimes
\left( \begin{array}{c} E \\
 D \end{array} \right) -
\left( \begin{array}{c} E \\
 D \end{array} \right) \otimes
\left( \begin{array}{c} \bar{E} \\
 \bar{D} \end{array} \right), \\ \\
<W| \left( \begin{array}{c} \bar{E} \\
 \bar{D} \end{array} \right)=0, \quad \left(\begin{array}{c} \bar{E} \\
 \bar{D} \end{array} \right) |V>=0
\end{array}
\end{equation}
in which the operators $\bar{E}$ and $\bar{D}$ are two auxiliary operators and $h$ is of the form (\ref{H2}). Using (\ref{MPA}) one can simply calculate the corresponding quadratic algebra of our model:
\begin{equation}
\begin{array}{l}
\label{FinalAlgebra}
[C,\bar{C}] = [E,\bar{E}] = 0, \\ \\
(\omega_{32}+\omega_{42}+\omega_{24}) EC - (\omega_{32}+\omega_{42}+\omega_{24}-\omega_{23})  E^{2} -
\omega_{24} C^{2}+\\(\omega_{24}-\omega_{23})CE=\bar{E}C-E\bar{C}, \\ \\
(\omega_{34}-\omega_{32})EC -(\omega_{23}+\omega_{43}+\omega_{34}-\omega_{32})E^{2} - \omega_{34}C^{2}+\\ (\omega_{34}+\omega_{23}+\omega_{43})CE= \bar{C}E-C\bar{E}, \\ \\
\langle W \vert\bar{E}=\langle W \vert \bar {C}=0, \\ \\
\bar{E} \vert V \rangle =\bar{C}\vert V \rangle=0
\end{array}
\end{equation}
in which we have defined $C:=D+E$ and $\bar{C}:=\bar{D}+\bar{E}$.\\
For the special tuning of the parameters given in (\ref{R2}) the authors in \cite{HSP} have shown, for the first time, that the quadratic algebra (\ref{FinalAlgebra}) has a four-dimensional matrix representation with non-diagonal $\bar{E}$ and $\bar{C}$. In our generalized model it turns out that the quadratic algebra (\ref{FinalAlgebra}) also has a four-dimensional matrix representation provided that the constraints (\ref{Con}) are fulfilled. Note that these constraints are automatically fulfilled for (\ref{R2}).\\
As long as the eigenvalues of the matrix $C$ are not equal and in an appropriate basis this matrix can be brought into a diagonal form. We have found that the matrices $C$ and $E$ besides the auxiliary matrices $\bar{C}$ and $\bar{E}$ have the following four-dimensional representation:
\begin{equation}
\label{DiagRepBulk}
\begin{array}{l}
C=\left(\begin{array}{cccc}
\frac{\delta_{1r}}{\delta_{1l}}\frac{\delta_{2r}}{\delta_{2l}}&0&0&0\\
0&(1-\rho)\frac{\delta_{2r}}{\delta_{2l}}&0&0\\
0&0&\frac{\delta_{2r}}{\delta_{2l}}&0\\
0&0&0&1\end{array} \right),\\
\bar{C}=\left(\begin{array}{cccc}
0&0&0&0\\
0&\rho\frac{\delta_{2r}}{\delta_{2l}}(\delta_{1r}-(1-\rho)\delta_{1l})&0&0\\
0&0&0&0\\
0&0&0&0
\end{array} \right),\\
E=\left(\begin{array}{cccc}
\frac{\delta_{1r}}{\delta_{1l}}\frac{\delta_{2r}}{\delta_{2l}}&e_{12}&e_{13}&e_{14}\\
0&(1-\rho)^2\frac{\delta_{2r}}{\delta_{2l}}&0&e_{24}\\
0&0&(1-\rho)\frac{\delta_{2r}}{\delta_{2l}}&e_{34}\\
0&0&0&1
\end{array} \right),\\
\bar{E}=\left(\begin{array}{cccc}
0&0&\bar{e}_{13}&\bar{e}_{14}\\
0&0&0&0\\
0&0&-\rho\frac{\delta_{2r}}{\delta_{2l}}(\delta_{1r}-(1-\rho)\delta_{1l})&\bar{e}_{34}\\
0&0&0&0
\end{array} \right)
\end{array}
\end{equation}
in which $e_{12}, e_{13}, \bar{e}_{13}, \bar{e}_{14}$ and $\bar{e}_{34}$ are functions of $\omega_{ij}$'s and are presented in the Appendix A. In this representation we have chosen the other parameters $e_{14}, e_{24}$ and $e_{34}$ in (\ref{DiagRepBulk}) as free parameters. On the other hand, the vectors $\vert V \rangle$ and $\langle W \vert$ have the following matrix representations:
\begin{equation}
\label{DiagRepBoun}
\vert V \rangle=\left(\begin{array}{c} v_1\\0\\1\\v_4
\end{array} \right), \quad \langle W \vert=\Bigl( 1,0,w_3,w_4 \Bigr)
\end{equation}
in which $v_1$ and $w_4$ are free parameters, while $v_4$ and $w_3$ are complicated functions of $\omega_{ij}$'s and are presented in the Appendix A. Nevertheless, if one looks for a nontrivial steady-state then the relation between $v_1$ and $w_4$  should be obtained by requiring that the probability of finding an empty lattice in long time limit is zero i.e.:
\begin{equation}
\label{Triv}
\langle W \vert E^L \vert V \rangle=0.
\end{equation}
By considering the constraint (\ref{Triv}) the partition function of the model can be easily calculated as follow:
\begin{equation}
\label{Partition1}
Z_L=a_1(\frac{\delta_{1r}}{\delta_{1l}}\frac{\delta_{2r}}{\delta_{2l}})^L+a_2((1-\rho)^2 \frac{\delta_{2r}}{\delta_{2l}})^L+a_3(\frac{\delta_{2r}}{\delta_{2l}})^L+a_4
\end{equation}
in which the coefficients $a_i$'s are function of $\omega_{ij}$'s. The explicit form of these coefficients are presented in the Appendix B. The phase diagram of the model can now be obtained by analyzing the thermodynamical behavior of the partition function $Z_L$ (i.e. in the limit $L\rightarrow\infty$) which obviously depends on the values of the hopping rates of the shock positions i.e. $\delta_{1l}, \delta_{1r}, \delta_{2l}$ and $\delta_{2r}$. One should note that the model has a mirror symmetry which means that it is invariant under the following transformations:
\begin{equation}
\label{SY}
\begin{array}{l}
\omega_{23} \leftrightarrow \omega_{32}\\
\omega_{24} \leftrightarrow \omega_{34}\\
\omega_{42} \leftrightarrow \omega_{43}\\
i\longrightarrow L-i+1.
\end{array}
\end{equation}
Defining two new variables $x:=\frac{\delta_{1r}}{\delta_{1l}}$ and $y:=\frac{\delta_{2r}}{\delta_{2l}}$, it can be shown using (\ref{Ref}) that the case ($x>1,y<1$) never happens because $\rho$ then has to be negative.
\begin{figure}
\centering
\includegraphics[height=5.5cm]{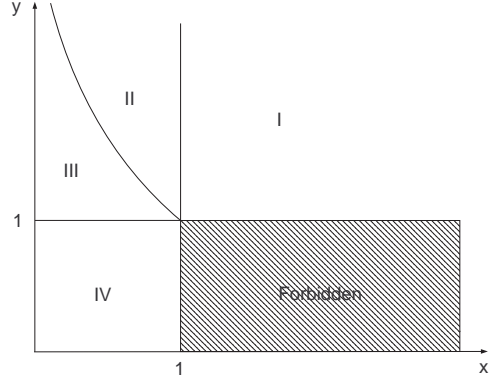}
\caption{The phase diagram of the model in terms of the two variables $x$ and $y$. The four regions I-IV are defined in the text. The coexistence line between the regions II and III is $y=\frac{1}{x}$. }
\end{figure}
In terms of the two variables $x$ and $y$ the phase diagram of the model has four regions defined as:
$$
\begin{array}{l}
\mbox{Region I for} \quad x>1,y>1,xy>1,\\
\mbox{Region II for} \quad x<1,y>1,xy>1,\\
\mbox{Region III for} \quad x<1,y>1,xy<1,\\
\mbox{Region IV for} \quad x<1,y<1,xy<1.
\end{array}
$$
The phase diagram of the model is shown in Figure 2. One can easily check that under the symmetry (\ref{SY}) the two variables $x$ and $y$ become $1/y$ and $1/x$ respectively. This implies that we need only to study two phases (either I and II or III and IV) since the information for the two other phases can be easily obtained using (\ref{SY}). In what follows we only consider the two phases III and IV. In the phase III the left random walker (the left shock front) has more tendency to move leftwards while the right random walker (the right shock front) has more tendency to move rightwards. In this phase the mean distance between the two random walkers is of order $L$ and therefore the density of particles in the bulk of the lattice increases (this can be understood by looking at the definition of $\vert P_{m,n} \rangle$). We call this phase the high-density phase. In contrast, in the phase IV both random walkers have more tendency to move leftwards, and therefore, the mean distance between the random walkers (the shock fronts) decreases. In this phase the mean distance between the two random walkers is of order $1$. Since the mean density of particles decreases in this phase, it is called the low-density phase. Since the matrix representation of the quadratic algebra is finite-dimensional, the density profile of the particles defined as:
\begin{equation}
\label{Prof}
\langle \rho_k \rangle=\frac{\sum_{\{\tau_i\}}\tau_k P({\tau})}{Z_L}=\frac{\langle W \vert C^{k-1}(C-E)C^{L-k}\vert V \rangle}{Z_L} \quad \mbox{for} \quad 1 \leq k \leq L
\end{equation}
either has linear or exponential behaviors. Note that in calculating (\ref{Prof}) the constraint (\ref{Triv}) should also be applied to its nominator. It turns out that (\ref{Prof}) has only exponential behaviors in both phases III and IV. In the low-density (high-density) phase the mean density of the particles in the bulk of the lattice is zero ($\rho$) while at the boundaries the density profile of the particles changes exponentially with two correlation lengths $\vert \ln y \vert^{-1}$ and $\vert \ln xy \vert^{-1}$ ($\vert \ln y \vert^{-1}$ and $\vert \ln x \vert^{-1}$).\\
It is also interesting to study the coexistence line between the two phases III and IV i.e. on the line $\frac{\delta_{2r}}{\delta_{2l}}=1$ for $\frac{\delta_{1r}}{\delta_{1l}}<1$. One should note that in this case two eigenvalues of the matrix $C$ become equal, and thereby cannot be diagonalized. On the coexistence line between the two phases III and IV, we have found the following matrix representation for the operators in (\ref{FinalAlgebra}):
\begin{equation}
\label{NonDiagRepBulk}
\begin{array}{l}
C=\left(\begin{array}{cccc}
\frac{\delta_{1r}}{\delta_{1l}}&0&0&0\\
0&(1-\rho)&0&0\\
0&0&1&c_{34}\\
0&0&0&1\end{array} \right),\\
\bar{C}=\left(\begin{array}{cccc}
0&0&0&0\\
0&\rho(\delta_{1r}-(1-\rho)\delta_{1l})&0&0\\
0&0&0&0\\
0&0&0&0
\end{array} \right),\\
E=\left(\begin{array}{cccc}
\frac{\delta_{1r}}{\delta_{1l}}&e_{12}&e_{13}&e_{14}\\
0&(1-\rho)^2&0&e_{24}\\
0&0&(1-\rho)&e_{34}\\
0&0&0&1
\end{array} \right), \\
\bar{E}=\left(\begin{array}{cccc}
0&0&\bar{e}_{13}&\bar{e}_{14}\\
0&0&0&0\\
0&0&-\rho(\delta_{1r}-(1-\rho)\delta_{1l})&\bar{e}_{34}\\
0&0&0&0
\end{array} \right)
\end{array}
\end{equation}
and for the vectors:
\begin{equation}
\label{NonDiagRepBoun}
\vert V \rangle=\left(\begin{array}{c} v_1\\0\\1\\v_4
\end{array} \right), \quad \langle W \vert=\Bigl( 1,0,w_3,w_4 \Bigr)
\end{equation}
which is valid if the reaction rates $\omega_{ij}$'s in (\ref{R1}) satisfy both (\ref{Con}) and the following constraint:
\begin{equation}
\omega_{43}=\omega_{42}(\frac{\omega_{24}+\omega_{34}}{\omega_{42}+\omega_{43}+\omega_{24}+\omega_{34}})^2
+\omega_{34}(\frac{\omega_{42}+\omega_{43}}{\omega_{42}+\omega_{43}+\omega_{24}+\omega_{34}})^2.
\end{equation}
As in the diagonal case some of the matrix elements are free. In (\ref{NonDiagRepBulk}) the free parameters are chosen to be $c_{34}, e_{14}, e_{24}$ and $e_{34}$. The rest of the matrix elements are functions of $\omega_{ij}$'s and are presented in the Appendix A. The two elements $v_1$ and $w_4$ can be eliminated from the calculations by applying the constraints (\ref{Triv}) while the other elements $v_4$ and $w_3$ are presented in the Appendix A. The partition function of the system, after applying the constraint (\ref{Triv}), is obtained to be:
\begin{equation}
\label{Partition2}
Z_L=b_1(\frac{\delta_{1r}}{\delta_{1l}})^L+b_2(1-\rho)^{2L}+b_3L+b_4
\end{equation}
where the coefficients $b_i$'s are complicated functions of the reaction rates $\omega_{ij}$'s. These coefficients are  presented in the Appendix B. As can be seen, on the coexistence line the left random walker (the left shock front) has more tendency to move leftwards while the right random walker (the right shock front) performs an unbiased random walk. In this case we expect that the density of particles increases near the left boundary. The density profile of the particles, defined in (\ref{Prof}), can be calculated on this line and it turns out that it is an exponentially increasing function near the left boundary with a characteristic length $\vert \ln x \vert^{-1}$ while it changes linearly in the bulk of the lattice.
%%%%%%%%%%%%%%%%%%%%%%%%%%%%%%%%%%%%%%%%%%%%%%%%%%%%%%%%%%%%%%%%%%%%%%%%%%%%%%%%%%%%%%%%%%%%%%%%%%%%%%%%%%%%%%%%%%%%%%%
\section{Summary and outlook}
In this paper we considered a generalized coagulation-decoagulation model on a finite lattice with reflecting boundaries and studied the conditions under which the steady-state probability distribution of the system can be written as a product of non-commutative operators. We started with a product shock measure with two shock fronts and obtained its dynamics generated by the Hamiltonian of the system. We found the necessary conditions on the microscopic reaction rates under which each of the shock fronts has a simple biased random walk dynamics. Finally by studying the steady-state of this system using the matrix-product approach we showed that its quadratic algebra has a four-dimensional matrix representation under the same conditions. The phase diagram of the model and also the density profile of the particles in each phase and on the coexistence line were studied. We saw that the whole phase diagram of the model can be described by the hopping rates of the shock fronts. The system has three different length scales. These are given by the eigenvalues of the matrix $C$ which can be written in terms of the hopping rates of the shock fronts. \\
One can easily check that for the special tuning of the parameters (\ref{R2}) the constraints (\ref{Con}) are satisfied and that all of our calculations reduce to those obtained in \cite{HSP}. It is interesting to investigate the solutions of (\ref{TE}) in long time limit and construct the steady-state of the system as a linear superposition of such solutions and then compare the results with those obtained from the matrix-product approach in present work. The results will be presented elsewhere.
%%%%%%%%%%%%%%%%%%%%%%%%%%%%%%%%%%%%%%%%%%%%%%%%%%%%%%%%%%%%%%%%%%%%%%%%%%%%%%%%%%%%%%%%%%%%%%%%%%%%%%%%%%%%%%%%%%%%%%%
\appendix
\section{The matrix elements}
Here we present the matrix elements for the diagonal matrices (\ref{DiagRepBulk}) and the vectors (\ref{DiagRepBoun}). The matrix elements of $E$ are obtained to be:
$$
\begin{array}{l}
e_{12}=((e_{14} (\omega _{24}+\omega _{34})^2 \omega _{43} (\omega _{24}^2 (\omega _{42}-\omega _{43})+\omega _{34}^2 (\omega _{42}-\omega _{43})+2 \omega _{34} \omega _{42} (\omega _{42}+\omega _{43})+\\\omega _{42} (\omega _{42}+\omega _{43})^2+\omega _{24} (\omega _{42}-\omega _{43}) (2 \omega _{34}+\omega _{42}+\omega _{43})) (\omega _{24}^2(\omega _{42}-\omega _{43})+\omega _{34}^2 (\omega _{42}-\omega _{43})-\\\omega _{43} (\omega _{42}+\omega _{43})^2+\omega _{34} (\omega _{42}^2-\omega _{43}^2)+2 \omega _{24} (\omega _{34} (\omega _{42}-\omega _{43})-\\ \omega _{43} (\omega _{42}+\omega _{43}))))/(e_{24} (\omega _{24}^2 \omega _{42}+ 2 \omega _{24} \omega _{34} \omega _{42}+\omega _{34} (\omega _{34} \omega _{42}+\\(\omega _{42}+\omega _{43})^2)) (\omega _{42} (\omega _{24} (-\omega _{42}+\omega _{43})+\omega _{43} (2 \omega _{34}+\omega _{42}+\omega _{43})) (\omega _{24}^2 \omega _{42}+\\2 \omega _{24} \omega _{34}\omega _{42}+\omega _{34} (\omega _{34} \omega _{42}+(\omega _{42}+\omega _{43})^2))+\omega _{43} (\omega _{42}
(2 \omega _{24}+\omega _{34}+\omega _{42})+(-\omega _{34}+\\\omega _{42}) \omega _{43}) (\omega _{24}^2 \omega _{43}+\omega_{34}^2 \omega _{43}+\omega _{24} (2 \omega _{34} \omega _{43}+(\omega _{42}+\omega _{43})^2))))),
\\ \\
e_{13}=((e_{14} \omega _{43} (\omega _{24}+\omega _{34}+\omega _{42}+\omega _{43})^2 (\omega _{24}^2 (\omega _{42}-\omega_{43})+\omega _{34} (\omega _{34} (\omega _{42}-\omega _{43})+\\ \omega _{42} (\omega _{42}+\omega _{43}))+\omega_{24} (2 \omega _{34} (\omega _{42}-\omega _{43})-\omega _{43} (\omega _{42}+\omega _{43})))^2)/(e_{34}(\omega _{42}+\omega _{43}) (\omega _{24}^2 \omega _{42}+\\2 \omega _{24} \omega _{34} \omega _{42}+\omega _{34} (\omega _{34}\omega _{42}+(\omega _{42}+\omega _{43})^2)) (\omega _{24}^3 \omega _{42} (\omega _{42}-2 \omega _{43})+\omega_{34} \omega _{43} (\omega _{34}^2 (-2 \omega _{42}+\\ \omega _{43})-3 \omega _{34} \omega _{42} (\omega _{42}+\omega _{43})- \omega_{42} (\omega _{42}+\omega _{43})^2)+\omega _{24}^2 (-3 \omega _{42} \omega _{43} (\omega _{42}+\omega _{43})+\\ \omega_{34} (2 \omega _{42}^2-6 \omega _{42} \omega _{43}+\omega _{43}^2))+ \omega _{24} (-\omega _{42} \omega _{43} (\omega_{42}+\omega _{43})^2+\\ \omega _{34} (\omega _{42}+\omega _{43}) (\omega _{42}^2-4 \omega _{42} \omega _{43}+\omega _{43}^2)+\omega_{34}^2 (\omega _{42}^2-6 \omega _{42} \omega _{43}+2 \omega _{43}^2))))).
\end{array}
$$
The other matrix elements $e_{14}$, $e_{24}$ and $e_{34}$ are free. For the matrix $\bar E$ we have found:
$$
\begin{array}{l}
\bar{e}_{13}=e_{13}\omega_{42},
\\ \\
\bar{e}_{14}=-(e_{14} \omega _{42} \omega _{43} (\omega _{24}+\omega _{34}+\omega _{42}+\omega _{43})^2 (\omega _{42} ((\omega _{24}+\omega _{34})^2+\omega _{34} \omega _{42})-\\((\omega _{24}+\omega _{34})^2+(\omega _{24}-\omega _{34}) \omega _{42}) \omega _{43}-\omega _{24} \omega _{43}^2))/((\omega _{42}+\omega _{43}) (\omega _{24} \omega _{42}^2 ((\omega _{24}+\omega _{34})^2+\\ \omega _{34} \omega _{42})-\omega _{42} (2 (\omega _{24}+\omega _{34})^3+3 (\omega _{24}^2+\omega _{24} \omega _{34}+\omega _{34}^2) \omega _{42}+(\omega _{24}+\omega _{34}) \omega _{42}^2) \omega _{43}+\\(\omega _{34} (\omega _{24}+\omega _{34})^2-3 (\omega _{24}^2+\omega _{24} \omega _{34}+\omega _{34}^2) \omega _{42}-2 (\omega _{24}+\omega _{34}) \omega _{42}^2) \omega _{43}^2+(\omega _{24} \omega _{34}-\\(\omega _{24}+\omega _{34}) \omega _{42}) \omega _{43}^3)),
\\ \\
\bar{e}_{34}=-e_{34} \omega _{43}.
\end{array}
$$
For the vector elements in (\ref{DiagRepBoun}) we have found:
$$
\begin{array}{l}
v_4=(\omega _{24}^2 (\omega _{42}-\omega _{43})+\omega _{34} (\omega _{34} (\omega _{42}-\omega _{43})+\omega
_{42} (\omega _{42}+\omega _{43}))+\\ \omega _{24} (2 \omega _{34} (\omega _{42}-\omega _{43})-\omega _{43} (\omega
_{42}+\omega _{43})))/(e_{34} (\omega _{24}^2 \omega _{42}+2 \omega _{24} \omega _{34} \omega _{42}+\\
\omega _{34} (\omega _{34}\omega _{42}+(\omega _{42}+\omega _{43})^2))),
\\ \\
w_3=-(e_{14} \omega _{42} (\omega _{24}+\omega _{34}+\omega _{42}+\omega _{43})^2 (\omega _{42} ((\omega _{24}+\omega _{34})^2+\omega _{34} \omega _{42})-\\((\omega _{24}+\omega _{34})^2+(\omega _{24}-\omega _{34}) \omega _{42}) \omega _{43}-\omega _{24} \omega _{43}^2))/(e_{34} (\omega _{42}+\omega _{43}) (\omega _{24} \omega _{42}^2 ((\omega _{24}+\omega _{34})^2+\\ \omega _{34} \omega _{42})-\omega _{42} (2 (\omega _{24}+\omega _{34})^3+3 (\omega _{24}^2+\omega _{24} \omega _{34}+\omega _{34}^2) \omega _{42}+(\omega _{24}+\omega _{34}) \omega _{42}^2) \omega _{43}+\\(\omega _{34} (\omega _{24}+\omega _{34})^2-3 (\omega _{24}^2+\omega _{24} \omega _{34}+\omega _{34}^2) \omega _{42}-2 (\omega _{24}+\omega _{34}) \omega _{42}^2) \omega _{43}^2+\\(\omega _{24} \omega _{34}-(\omega _{24}+\omega _{34}) \omega _{42}) \omega _{43}^3)).
\end{array}
$$
On the other hand for the non-diagonal matrices (\ref{NonDiagRepBulk}) and the vectors (\ref{NonDiagRepBoun}) we have found the following results. The matrix elements of $E$ are obtained to be:
$$
\begin{array}{l}
e_{12}=(- c_{34} e_{14} ({\omega _{42}^2}-{\omega _{43}^2})(\omega _{34}^2 (\omega _{42}-\omega _{43})^2+(\omega _{42}-\omega _{43})(- \omega _{34} \omega _{42}^2-6 \omega _{34} \omega _{42}\omega _{43}-\\ \omega _{43}^2\omega _{34})+ \omega _{43} \omega _{42}^3+\omega _{42} \omega _{43}^3+6 \omega _{42}^2 \omega _{43}^2+\sqrt{\omega _{43}(\omega _{34}+\omega _{42})-\omega _{34} \omega _{42}}(-2\omega _{34}(\omega _{42}^2-\omega _{43}^2)+\\4\omega _{43} \omega _{42}(\omega _{42}+\omega _{43}))))/(e_{24} (\sqrt{\omega _{43}(\omega _{34}+\omega _{42})-\omega _{34} \omega _{42}}(e_{34} \omega _{43} (\omega _{42}^2 (\omega _{34}^2-10 \omega _{34} \omega _{42}+\\5 \omega _{42}^2)-2 (\omega _{34}-5 \omega _{42}) \omega _{42} (\omega _{34}+\omega _{42}) \omega _{43}+(\omega _{34}+\omega _{42})^2 \omega _{43}^2)+c_{34} \omega _{42} (\omega _{34}^2 (\omega _{42}-\omega _{43})^2-\\2 \omega _{34} (\omega _{42}-\omega _{43}) (3 \omega _{42}^2+\omega _{42} \omega _{43}+2 \omega _{43}^2)+\omega _{42} (\omega _{42}^3+6 \omega _{42}^2 \omega _{43}+5 \omega _{42} \omega _{43}^2+4 \omega _{43}^3)))+\\e_{34} \omega _{42} \omega _{43} (\omega _{42}^2 (5 \omega _{34}^2-10 \omega _{34} \omega _{42}+\omega _{42}^2)+10 \omega _{42} (-\omega _{34}^2+\omega _{42}^2) \omega _{43}+5 (\omega _{34}+\omega _{42})^2 \omega _{43}^2)+\\c_{34} (\omega _{34}^2 (\omega _{42}-\omega _{43})^2 (4 \omega _{42}^2+\omega _{43}^2)-2 \omega _{34} \omega _{42} (\omega _{42}-\omega _{43}) (\omega _{42}+\omega _{43}) (2 \omega _{42}^2+2 \omega _{42} \omega _{43}+\\\omega _{43}^2)+\omega _{42}^2 \omega _{43} (4 \omega _{42}^3+5 \omega _{42}^2 \omega _{43}+6 \omega _{42} \omega _{43}^2+\omega _{43}^3)))),
\\ \\
e_{13}=(-e_{14} \omega _{43} (\omega _{42}-\omega _{43}))/((c_{34}\omega _{42}+e_{34} \omega _{43})\sqrt{\omega _{43}(\omega _{34}+\omega _{42})-\omega _{34} \omega _{42}}+\\ e_{34} \omega _{43}\omega _{42}+c_{34}\omega _{43}^2).
\end{array}
$$
As in the diagonal case the other matrix elements $e_{14}$, $e_{24}$ and $e_{34}$ are free. The matrix elements for $\bar E$ are found to be:
$$
\begin{array}{l}
\bar{e}_{13}=e_{13}\omega_{42},\\ \\
\bar{e}_{14}=((e_{14}e_{34}\omega _{42}(\omega _{34}+\omega _{42})\omega _{43}))/(c_{34}(\omega _{43}+\omega _{42})\sqrt{\omega _{43}(\omega _{34}+\omega _{42})-\omega _{34} \omega _{42}}+\\e_{34}(\omega _{34}+\omega _{42})\omega _{43}+c_{34}\omega _{42}(\omega _{34}-\omega _{43})),
\\ \\
\bar{e}_{34}=-e_{34} \omega _{43}.
\end{array}
$$
The non-free vector elements in this case are given by:
$$
\begin{array}{l}
v_4=(\omega _{42}-\omega _{43})/(e_{34} (\omega _{42}+\sqrt{\omega _{43}(\omega _{34}+\omega _{42})-\omega _{34} \omega _{42}}))\\ \\
w_3=(e_{14} \omega _{42} (\omega _{34}+\omega _{42}))/(e_{34} \omega _{43}(\omega _{34}+\omega _{42})+c_{34} (\omega _{42}(\omega _{34}-\omega_{43})+\\(\omega _{42}+\omega _{43})\sqrt{\omega _{43}(\omega _{34}+\omega _{42})-\omega _{34} \omega _{42}}))
\end{array}
$$
\section{The partition function}
The coefficients $a_i$'s for $i=1,\cdots,4$ in (\ref{Partition1}) are found to be:
$$
\begin{array}{l}
a_1=-(e_{14} \omega _{42} (\omega _{42} ((\omega _{24}+\omega _{34})^2+\omega _{34} \omega _{42})-((\omega _{24}+\omega _{34})^2+(\omega _{24}-\omega _{34}) \omega _{42}) \omega _{43}-\\ \omega _{24} \omega _{43}^2) (\omega _{42} ((\omega _{24}+\omega _{34})^2+\omega _{34} \omega _{42})-((\omega _{24}+\omega _{34})^2+2 \omega _{24} \omega _{42}+\omega _{42}^2) \omega _{43}-(2 \omega _{24}+\\ \omega _{34}+2 \omega _{42}) \omega _{43}^2- \omega _{43}^3) (\omega _{24}^2+\omega _{34} (\omega _{34}+\omega _{42})+\omega _{24} (2 (\omega _{34}+\omega _{42})+\omega _{43})))/(e_{34} (\omega _{42}+\\ \omega _{43}) (\omega _{42} (-(\omega _{24}+\omega _{34})^2+\omega _{24} \omega _{42})+((\omega _{24}+\omega _{34})^2+2 \omega _{24} \omega _{42}) \omega _{43}+\omega _{24} \omega _{43}^2) (\omega _{24} \omega _{42}^2 \\ ((\omega _{24}+\omega _{34})^2+\omega _{34} \omega _{42})-\omega _{42} (2 (\omega _{24}+\omega _{34})^3+3 (\omega _{24}^2+\omega _{24} \omega _{34}+\omega _{34}^2) \omega _{42}+(\omega _{24}+\\ \omega _{34}) \omega _{42}^2) \omega _{43}+(\omega _{34} (\omega _{24}+\omega _{34})^2-3 (\omega _{24}^2+\omega _{24} \omega _{34}+\omega _{34}^2) \omega _{42}-2 (\omega _{24}+\omega _{34}) \omega _{42}^2) \omega _{43}^2\\+(\omega _{24} \omega _{34}-(\omega _{24}+\omega _{34}) \omega _{42}) \omega _{43}^3)),
\end{array}
$$

$$
\begin{array}{l}
a_2=-(e_{14} (\omega _{24}+\omega _{34})^2 \omega _{42} (\omega _{42} ((\omega _{24}+\omega _{34})^2+\omega _{34} \omega _{42})-((\omega _{24}+\omega _{34})^2+\\(\omega _{24}-\omega _{34}) \omega _{42}) \omega _{43}-\omega _{24} \omega _{43}^2) (\omega _{42} (\omega _{24}^2+(\omega _{34}+\omega _{42})^2+\omega _{24} (2 \omega _{34}+\omega _{42}))-\\((\omega _{24}+\omega _{34})^2-2 \omega _{34} \omega _{42}-2 \omega _{42}^2) \omega _{43}-(\omega _{24}-\omega _{42}) \omega _{43}^2) (\omega _{42} ((\omega _{24}+\omega _{34})^2+\omega _{34} \omega _{42})-\\((\omega _{24}+\omega _{34})^2+2 \omega _{24} \omega _{42}+\omega _{42}^2) \omega _{43}-(2 \omega _{24}+\omega _{34}+2 \omega _{42}) \omega _{43}^2-\omega _{43}^3))/(e_{34} (\omega _{42}+\omega _{43}) \\(\omega _{42} (-(\omega _{24}+\omega _{34})^2+\omega _{24} \omega _{42})+((\omega _{24}+\omega _{34})^2+2 \omega _{24} \omega _{42}) \omega _{43}+\omega _{24} \omega _{43}^2) \\(\omega _{42} ((\omega _{24}+\omega _{34})^2+\omega _{34} \omega _{42})-((\omega _{24}+\omega _{34})^2-2 \omega _{34} \omega _{42}) \omega _{43}+\omega _{34} \omega _{43}^2)\\ (\omega _{24} \omega _{42}^2 (-(\omega _{24}+\omega _{34})^2-\omega _{34} \omega _{42})+\omega _{42} (2 (\omega _{24}+\omega _{34})^3+\\3 (\omega _{24}^2+\omega _{24} \omega _{34}+\omega _{34}^2) \omega _{42}+(\omega _{24}+\omega _{34}) \omega _{42}^2) \omega _{43}+(-\omega _{34} (\omega _{24}+\omega _{34})^2+\\3 (\omega _{24}^2+\omega _{24} \omega _{34}+\omega _{34}^2) \omega _{42}+2 (\omega _{24}+\omega _{34}) \omega _{42}^2) \omega _{43}^2+(-\omega _{24} \omega _{34}+(\omega _{24}+\omega _{34}) \omega _{42}) \omega _{43}^3)),
\end{array}
$$

$$
\begin{array}{l}
a_3=-(e_{14} \omega _{42} (\omega _{24}+\omega _{34}+\omega _{42}+\omega _{43})^2 (\omega _{42} ((\omega _{24}+\omega _{34})^2+\omega _{34} \omega _{42})-((\omega _{24}+\omega _{34})^2+\\(\omega _{24}-\omega _{34}) \omega _{42}) \omega _{43}-\omega _{24} \omega _{43}^2))/(e_{34} (\omega _{42}+\omega _{43}) (\omega _{24} \omega _{42}^2 ((\omega _{24}+\omega _{34})^2+\omega _{34} \omega _{42})-\\ \omega _{42} (2 (\omega _{24}+\omega _{34})^3+3 (\omega _{24}^2+\omega _{24} \omega _{34}+\omega _{34}^2) \omega _{42}+(\omega _{24}+\omega _{34}) \omega _{42}^2) \omega _{43}+(\omega _{34} (\omega _{24}+\\ \omega _{34})^2-3 (\omega _{24}^2+\omega _{24} \omega _{34}+\omega _{34}^2) \omega _{42}-2 (\omega _{24}+\omega _{34}) \omega _{42}^2) \omega _{43}^2+(\omega _{24} \omega _{34}-(\omega _{24}+\\ \omega _{34}) \omega _{42}) \omega _{43}^3)),
\end{array}
$$

$$
\begin{array}{l}
a_4=(e_{14} \omega _{42} (\omega _{42} ((\omega _{24}+\omega _{34})^2+\omega _{34} \omega _{42})-((\omega _{24}+\omega _{34})^2+(\omega _{24}-\omega _{34}) \omega _{42}) \omega _{43}-\\ \omega _{24} \omega _{43}^2) (\omega _{42} (\omega _{24}^2+(\omega _{34}+\omega _{42})^2+\omega _{24} (2 \omega _{34}+\omega _{42}))-((\omega _{24}+\omega _{34})^2-2 \omega _{34} \omega _{42}-\\ 2 \omega _{42}^2) \omega _{43}-(\omega _{24}-\omega _{42}) \omega _{43}^2) (\omega _{24}^2+\omega _{24} (2 \omega _{34}+\omega _{43})+\omega _{34} (\omega _{34}+\omega _{42}+\\2 \omega _{43})))/(e_{34} (\omega _{42}+\omega _{43}) (\omega _{42} ((\omega _{24}+\omega _{34})^2+\omega _{34} \omega _{42})-((\omega _{24}+\omega _{34})^2-\\2 \omega _{34} \omega _{42}) \omega _{43}+\omega _{34} \omega _{43}^2) (\omega _{24} \omega _{42}^2 ((\omega _{24}+\omega _{34})^2+\omega _{34} \omega _{42})-\omega _{42} (2 (\omega _{24}+\omega _{34})^3+\\3 (\omega _{24}^2+\omega _{24} \omega _{34}+\omega _{34}^2) \omega _{42}+(\omega _{24}+\omega _{34}) \omega _{42}^2) \omega _{43}+(\omega _{34} (\omega _{24}+\omega _{34})^2-\\3 (\omega _{24}^2+\omega _{24} \omega _{34}+\omega _{34}^2) \omega _{42}-2 (\omega _{24}+\omega _{34}) \omega _{42}^2) \omega _{43}^2+(\omega _{24} \omega _{34}-(\omega _{24}+\omega _{34}) \omega _{42}) \omega _{43}^3)),
\end{array}
$$
The coefficients $b_i$'s for $i=1,\cdots,4$ in (\ref{Partition2}) are also found to be:

$$
\begin{array}{l}
b_1=(-c_{34} e_{14} \omega _{42} \omega _{43}(\omega _{42}^2 \omega _{43} (\omega _{42}+\omega _{43}) (3 \omega _{42}+\omega _{43})^2+\omega _{34}^2 (\omega _{42}-\omega _{43})^2 (8 \omega _{42}^2+\\ \omega _{42} \omega _{43}+\omega _{43}^2)-2 \omega _{34} \omega _{42} (\omega _{42}-\omega _{43}) (4 \omega _{42}^3+11 \omega _{42}^2 \omega _{43}+4 \omega _{42} \omega _{43}^2+\omega _{43}^3)+\\\sqrt{\omega _{43}(\omega _{34}+\omega _{42})-\omega _{34} \omega _{42}}(2 \omega _{42} (\omega _{34}^2 (\omega _{42}-\omega _{43})^2+2 \omega _{34} (-3 \omega _{42}^3+\omega _{42}^2 \omega _{43}+\omega _{42} \omega _{43}^2+\\ \omega _{43}^3)+\omega _{42} (\omega _{42}^3+8 \omega _{42}^2 \omega _{43}+5 \omega _{42} \omega _{43}^2+2 \omega _{43}^3)))))/(e_{34} (\omega _{42}^2-\omega _{43}^2)((-\omega _{34} \omega _{42}+(\omega _{34}+\\ \omega _{42}) \omega _{43}) (4 e_{34} \omega _{42} \omega _{43} (\omega _{34} (-\omega _{42}+\omega _{43})+\omega _{42} (\omega _{42}+\omega _{43}))+c_{34} (\omega _{42} (\omega _{42}+\omega _{43})^3\\-\omega _{34} (\omega _{42}-\omega _{43}) (3 \omega _{42}^2+\omega _{43}^2)))+\sqrt{\omega _{43}(\omega _{34}+\omega _{42})-\omega _{34} \omega _{42}}(e_{34} \omega _{43} (\omega _{42}^2 (\omega _{34}^2-\\6 \omega _{34} \omega _{42}+\omega _{42}^2)-2 (\omega _{34}-3 \omega _{42}) \omega _{42} (\omega _{34}+\omega _{42}) \omega _{43}+(\omega _{34}+\omega _{42})^2 \omega _{43}^2)+\\ c_{34} \omega _{42} (\omega _{34}^2 (\omega _{42}-\omega _{43})^2-\omega _{34} (\omega _{42}-\omega _{43}) (3 \omega _{42}^2+2 \omega _{42} \omega _{43}+3 \omega _{43}^2)+\\ \omega _{42} \omega _{43} (3 \omega _{42}^2+2 \omega _{42} \omega _{43}+3 \omega _{43}^2))))),
\end{array}
$$

$$
\begin{array}{l}
b_2=(c_{34} e_{14} \omega _{42}(\omega _{42}+\omega _{43})(\omega _{34}^2 (\omega _{42}-\omega _{43})^2 (3 \omega _{42}+2 \omega _{43})+\omega _{42}^2 \omega _{43} (\omega _{42}^2+10 \omega _{42} \omega _{43}+\\ 5 \omega _{43}^2)- \omega _{34} \omega _{42} (\omega _{42}-\omega _{43}) (\omega _{42}^2+12 \omega _{42} \omega _{43}+7 \omega _{43}^2)+\sqrt{\omega _{43}(\omega _{34}+\omega _{42})-\omega _{34} \omega _{42}}(\omega _{34}^2 \\(\omega _{42}-\omega _{43})^2- \omega _{34} (\omega _{42}-\omega _{43}) (3 \omega _{42}^2+8 \omega _{42} \omega _{43}+\omega _{43}^2)+\omega _{42} \omega _{43} (5 \omega _{42}^2+10 \omega _{42} \omega _{43}+\\ \omega _{43}^2))))/(e_{34} (\omega _{42}-\omega _{43})((-\omega _{34} \omega _{42}+(\omega _{34}+\omega _{42}) \omega _{43}) (-c_{34} (5 \omega _{34}-\omega _{42}) \omega _{42}^3+\\ \omega _{42}^2 (e_{34} (-7 \omega _{34}+5 \omega _{42})+c_{34} (4 \omega _{34}+6 \omega _{42})) \omega _{43}+\omega _{42} (-(c_{34}-6 e_{34}) \omega _{34}+\\ 5 (c_{34}+2 e_{34}) \omega _{42}) \omega _{43}^2+(e_{34} (\omega _{34}+\omega _{42})+2 c_{34} (\omega _{34}+2 \omega _{42})) \omega _{43}^3)+\\ \sqrt{\omega _{43}(\omega _{34}+\omega _{42})-\omega _{34} \omega _{42}}(e_{34} \omega _{43} (2 \omega _{34}^2 (\omega _{42}-\omega _{43})^2- \\ \omega _{34} \omega _{42} (\omega _{42}-\omega _{43}) (9 \omega _{42}+7 \omega _{43})+\omega _{42}^2 (\omega _{42}^2+\\ 10 \omega _{42} \omega _{43}+5 \omega _{43}^2))+ c_{34} (2 \omega _{34}^2 \omega _{42} (\omega _{42}-\omega _{43})^2-\\ \omega _{34} (\omega _{42}-\omega _{43}) (4 \omega _{42}^3+6 \omega _{42}^2 \omega _{43}+5 \omega _{42} \omega _{43}^2+\omega _{43}^3)+\\ \omega _{42} \omega _{43} (4 \omega _{42}^3+5 \omega _{42}^2 \omega _{43}+6 \omega _{42} \omega _{43}^2+\omega _{43}^3))))),
\end{array}
$$

$$
\begin{array}{l}
b_3=(c_{34} e_{14} \omega _{42} (\omega _{42}-\omega _{43}))/(e_{34} (e_{34} \omega _{43}+c_{34}\omega _{42} )\sqrt{\omega _{43}(\omega _{34}+\omega _{42})-\omega _{34} \omega _{42}}+\\e_{34}\omega _{43}( e_{34} \omega _{42}+c_{34}\omega _{43})),
\end{array}
$$

$$
\begin{array}{l}
b_4=(-c_{34} e_{14} \omega _{42} (\omega _{42}-\omega _{43})(4 \omega _{42} (\omega _{34} \omega _{42}-(\omega _{34}+\omega _{42}) \omega _{43}) (\omega _{34} (\omega _{42}-\omega _{43})-\\ \omega _{42} (\omega _{42}+\omega _{43}))+ \sqrt{\omega _{43}(\omega _{34}+\omega _{42})-\omega _{34} \omega _{42}}(\omega _{42}^2 (\omega _{34}^2-6 \omega _{34} \omega _{42}+\omega _{42}^2)-\\ 2 (\omega _{34}-3 \omega _{42}) \omega _{42} (\omega _{34}+\omega _{42}) \omega _{43}+(\omega _{34}+\omega _{42})^2 \omega _{43}^2)))/(e_{34} (\omega _{42}+\omega _{43})((3 \omega _{42}+\omega _{43})\\ (-\omega _{34} \omega _{42}+ (\omega _{34}+\omega _{42}) \omega _{43}) (c_{34} \omega _{42} (\omega _{42} (-\omega _{34}+\omega _{42})+(\omega _{34}+\omega _{42}) \omega _{43}+2 \omega _{43}^2)+\\e_{34} \omega _{43} (\omega _{34} (-\omega _{42}+\omega _{43})+\omega _{42} (3 \omega _{42}+\omega _{43})))+(-2 \omega _{34} (\omega _{42}-\omega _{43})+\\ \omega _{42} (\omega _{42}+3 \omega _{43})) (e_{34} \omega _{42} \omega _{43} (\omega _{42} (-3 \omega _{34}+\omega _{42})+3 (\omega _{34}+\omega _{42}) \omega _{43})+\\ c_{34} (-\omega _{34} (\omega _{42}-\omega _{43}) (2 \omega _{42}^2+\omega _{43}^2)+\omega _{42} \omega _{43} (2 \omega _{42}^2+\omega _{42} \omega _{43}+\omega _{43}^2)))+\\ \sqrt{\omega _{43}(\omega _{34}+\omega _{42})-\omega _{34} \omega _{42}}(2 e_{34} \omega _{43} (\omega _{34}^2 (\omega _{42}-\omega _{43})^2+\omega _{42}^2 (3 \omega _{42}+\\ \omega _{43}) (\omega _{42}+3 \omega _{43})+4 \omega _{34} \omega _{42} (-2 \omega _{42}^2+\omega _{42} \omega _{43}+\omega _{43}^2))+\\ c_{34} (2 \omega _{34}^2 \omega _{42} (\omega _{42}-\omega _{43})^2+\omega _{34} (-9 \omega _{42}^4+2 \omega _{42}^3 \omega _{43}+\\ 6 \omega _{42} \omega _{43}^3+\omega _{43}^4)+\omega _{42} (\omega _{42}^4+10 \omega _{42}^3 \omega _{43}+\\ 10 \omega _{42}^2 \omega _{43}^2+10 \omega _{42} \omega _{43}^3+\omega _{43}^4))))).
\end{array}
$$
\section*{Acknowledgements}
FHJ would like to acknowledge the financial support provided by Bu-Ali Sina University.

\end{document}